%% file: main.tex
\documentclass[a4paper,twoside]{article}

\usepackage{epsfig}
\usepackage{subcaption}
\usepackage{calc}
\usepackage{amssymb}
\usepackage{amstext}
\usepackage{amsmath}
\usepackage{amsthm}
\usepackage{multicol}
\usepackage{pslatex}
\usepackage{apalike}
\usepackage{multirow}
\usepackage[normalem]{ulem}
\useunder{\uline}{\ul}{}
\usepackage[table,xcdraw]{xcolor}
\usepackage{SCITEPRESS}     

\begin{document}

\title{Acoustic Anomaly Detection for Machine Sounds based on Image Transfer Learning}

\author{\authorname{Robert Müller\orcidAuthor{0000-0003-3108-713X}, Fabian Ritz\orcidAuthor{0000-0001-7707-1358}, Steffen Illium\orcidAuthor{0000-0003-0021-436X} and Claudia Linnhoff-Popien\orcidAuthor{0000-0001-6284-9286}}
\affiliation{Mobile and Distributed Systems Group, LMU Munich, Germany}
\email{\{robert.mueller, fabian.ritz, steffen.illium, linnhoff\}@ifi.lmu.de}
}


\input{sections/abstract}

\onecolumn \maketitle \normalsize \setcounter{footnote}{0} \vfill
\input{sections/introduction}
\input{sections/related_work}
\input{sections/proposed_system}
\input{sections/dataset}
\input{sections/experiments}
\input{sections/results}
\input{sections/conclusion}

\bibliographystyle{apalike}
{\small
\bibliography{mybib}}

\end{document}

%% file: sections/abstract.tex
\abstract{In industrial applications, the early detection of malfunctioning factory machinery is crucial. In this paper, we consider acoustic malfunction detection via transfer learning. Contrary to the majority of current approaches which are based on deep autoencoders, we propose to extract features using neural networks that were pretrained on the task of image classification. We then use these features to train a variety of anomaly detection models and show that this improves results compared to convolutional autoencoders in recordings of four different factory machines in noisy environments. Moreover, we find that features extracted from ResNet based networks yield better results than those from AlexNet and Squeezenet. In our setting, Gaussian Mixture Models and One-Class Support Vector Machines achieve the best anomaly detection performance.}

%% file: sections/introduction.tex
\section{Introduction}
Anomaly detection is one of the most prominent industrial applications of machine learning. It is used for video surveillance, monitoring of critical infrastructure or the detection of fraudulent behavior. However, most of the current approaches are based on detecting anomalies in the visual domain.
Issues arise when the scenery cannot be covered by cameras completely, leading to blind-spots in which no prediction can be made. Naturally, this applies to many internals of industrial production facilities and machines. In many cases a visual inspection can not capture the true condition of the surveilled entity. A pump suffering from a small leakage, a slide rail that has no grease or a fan undergoing voltage changes might appear intact when inspected visually but when monitored acoustically, reveal its actual condition through distinct sound patterns. Further, acoustic monitoring has the advantage of comparably cheap and easily deployable hardware. The early detection of malfunctioning machinery with a reliable acoustic anomaly detection system can prevent greater damages and reduce repair and maintenance costs.\\
\begin{figure}
  \centering
  \includegraphics[width=0.65\linewidth]{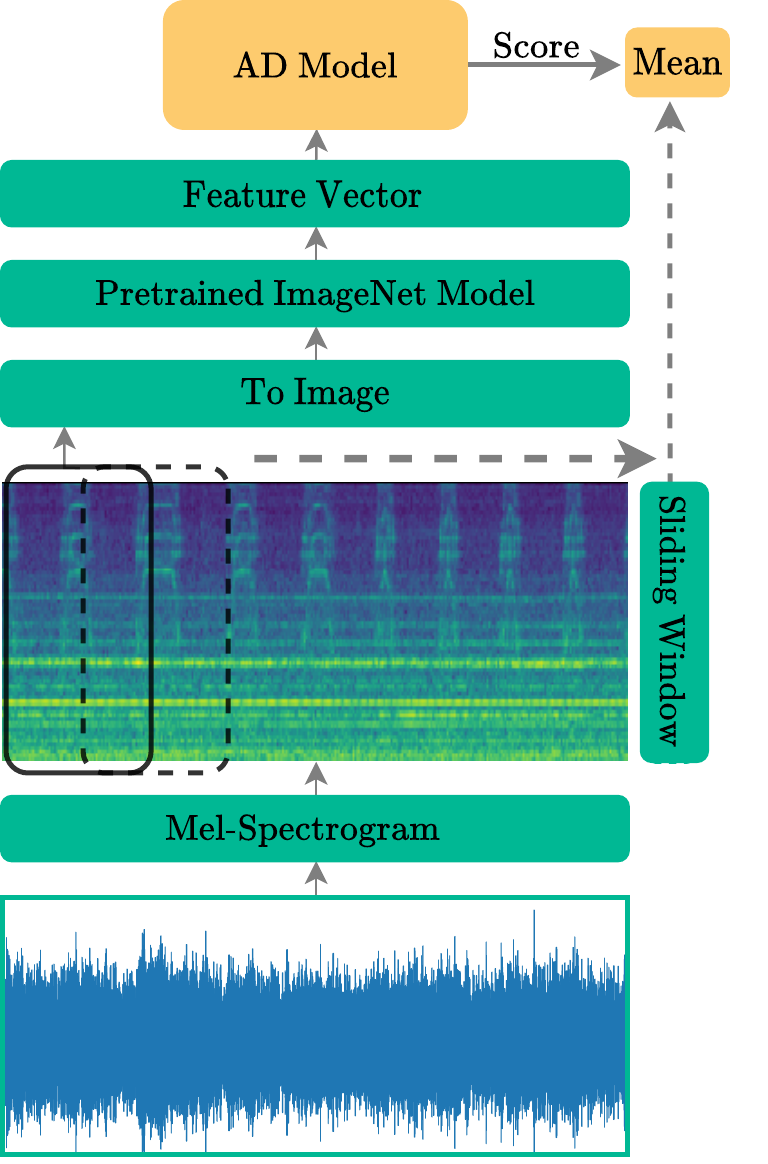}
  \caption{Overview of the proposed workflow. First, the raw waveform is transformed into a Mel-spectrogram. Small segments of $\approx 2s$ are then extracted in sliding window fashion. Subsequently, a pretrained image classification neural network is used to extract feature vectors. These feature vectors serve as the input to an anomaly detection model. A prediction over the whole recording is made by mean-pooling the scores of the analyzed segments.}
  \label{fig:model}
\end{figure}
In this work, we focus on the detection of anomalous sounds emitted from factory machinery such as fans, pumps, valves and slide rails. Obtaining an exhaustive number of recordings from anomalous operation for training is not suitable as it would require either deliberately damaging machines or waiting a potentially long time until enough machines suffered from damages. Consequently, we assume there is no access to anomalous recordings during the training of the anomaly detection systems. Hence, training the system proceeds in a fully unsupervised manner. Moreover, we assume normal operation recordings to be highly contaminated with background noises from real world factory environments.\\
In the recent years, using CNNs in conjunction with a signals time-frequency representations has become ubiquitous in acoustic signal processing for a variety of tasks such as environmental sound classification~\cite{salamon2017}, speech recognition~\cite{qian2016} and music audio tagging~\cite{Pons2019}. Nevertheless, these approaches specifically design CNN architectures for the task at hand and require a labeled dataset. These results make evident that CNNs are promising candidates for acoustic anomaly detection. Due to the lack of labels the predominant approach is to rely on deep autoencoders (AEs). An AE is a neural network (NN) that first compresses its input into a low dimensional representation and subsequently reconstructs the input. The reconstruction error is taken as the anomaly score since it is assumed that input differing from the training data cannot be reconstructed precisely. These is different to the more traditional approach where one extracts a set of handcrafted features (requires domain knowledge) from the signal and use these features as input to a dedicated anomaly detection (AD) model e.g. a density estimator. However, these AD models collapse with high dimensional input (e.g. images or spetrograms) due to the curse of deminsionality.\\
In this work we aim to combine the best of both worlds and ask the question whether it is possible to use a NN to automatically extract features and use these features in conjunction with more traditional anomaly detection models while  achieving comparable or even superior performance. 
By observing that patterns of anomalous operation can often be spotted visually in the time-frequency representation (e.g. Mel-spectrogram) of a recording, we claim that pretrained image classification convolutional neural networks (CNNs) can extract useful features even though the task at hand is vastly different. This is because in order to correctly classify images the CNN has to learn a generic filters such as edge, texture and object detectors~\cite{Olah2017,Olah2020} that can extract valuable and semantically meaningful features that also transfer to various downstream tasks. Moreover, this reduces the burden of finding a suitable neural network architecture.\\
We propose to use features from images of segments gathered from the Mel-spectrograms of normal operation data. We then standardize the obtained features and use them to train various anomaly detection models. A sliding window in combination with mean-pooling is used to make a decision over a longer time horizon at test time. A visualization of the proposed system can be seen in Figure \ref{fig:model}.\\

The remaining paper is structured as follows:\\
In Section \ref{sec:related_work}, we survey related approaches to acoustic anomaly detection in an unsupervised learning learning setting. Section \ref{sec:proposed_approach} introduces the proposed approach with more mathematical rigor. Then we briefly introduce the dataset we used to evaluate our method in Section \ref{sec:dataset}, followed by a description of the experimental setup in Section \ref{sec:experiments}. Results are discussed in Section \ref{sec:results}. We close by summarizing our findings and outlining future work in Section \ref{sec:conclusion}.

%% file: sections/related_work.tex
\section{Related Work}
\label{sec:related_work}
While various approaches on classification~\cite{Mesaros2018,Abesser2020} and tagging~\cite{Fonseca2019} of acoustic scenes have been proposed in the last years, acoustic anomaly detection is still underrepresented. Due to the release of publicly available datasets~\cite{Jiang2018,Purohit2019,Koizumi2019,Grollmisch2019}, the situation is gradually improving.\\
As previously mentioned, the majority of approaches to acoustic anomaly detection relies upon deep autoencoders. For example, 
\cite{Marchi2015} use a bidirectional recurrent denoising AE to reconstruct auditory spectral features to detect novel events. \cite{Duman2019} propose to use a convolutional AE on Mel-spectrograms to detect anomalies in the context of industrial plants and processes. In~\cite{Meire2019}, the authors compare various AE architectures with special focus on the applicability of these methods on the edge. They conclude that a convolutional architecture operating on the Mel-Frequency Cesptral coefficients is well suited for the task while a One-Class Support Vector Machine represents a strong and more parameter efficient baseline. \cite{Kawaguchi2019} explicitly address the issue of background noise. An ensemble method of front-end modules and back-end modules followed by an ensemble-based detector combines the strengths of various algorithms. Front-ends consist of blind-dereverberation and anomalous-sound-extraction algorithms, back-ends are AEs. The final anomaly score is computed by score-averaging. Finally, in \cite{Koizumi2017} anomalous sound detection is interpreted as statistical hypothesis testing where they propose a loss function based on the Neyman-Pearson lemma. However this approach relies on the simulation of anomalous sounds using expensive rejection sampling.\\
In contrast to these architecture-driven approaches, \cite{Koizumi2019BU} introduced Batch-Uniformization, a modification to the AE's training-procedure where the reciprocal of the probabilistic density of each sample is used to up-weigh rare sounds.\\
Another line of work investigates upon methods that operate directly on the raw waveform~\cite{Hayashi2018,Rushe2019}. These methods use generative, WaveNet-like~\cite{Oord2016} architectures to predict the next sample and take the prediction error as a measure of abnormality. Their results indicate a slight advantage over AE based approaches at the cost of higher computational demands.
In this work, we propose a different approach to acoustic anomaly detection. We use features extracted from NNs pretrained with image classification to train anomaly detection models, which is inspired by the success of these features in other areas, such as snore sound classification~\cite{Amiriparian2017}, emotion recognition in speech~\cite{Cummins2017}, music information retrieval~\cite{Gwardys2014} and medical applications~\cite{Amiriparian2020}.

%% file: sections/proposed_system.tex
\section{Proposed Approach}
\label{sec:proposed_approach}
Let $X \in \mathbb{R}^{F \times T}$ be the time-frequency representation of some acoustic recording where $T$ is the time dimension and $F$ the number of frequency bins. In the context of acoustic anomaly detection, we want to find a function $\mathcal{F}: X \rightarrow \mathbb{R}$ such that $\mathcal{F}(X)$ is higher for anomalous recordings than for recordings from normal operation without having access to anomalous recordings during training.
To reduce computational demands and to increase the number of datapoints, it is common to extract smaller patches $x_1, \dots x_i, \dots x_n$  of the underlying spectrogram $X$ across the time dimension in a sliding window fashion where $x_i \in \mathbb{R}^{t \times F}, t < T$. Here we propose to extract a $d$-dimensional feature vector using a feature extractor $f: \mathbb{R}^{t \times F} \rightarrow \mathbb{R}^d$ for each $x_i$. Then we can set $\mathcal{F}$ to be some anomaly detection algorithm and train $\mathcal{F}$ on all features of extracted patches in the dataset $\mathcal{D}=\{X_j \in \mathbb{R}^{F \times T}\}_{j=1}^{N}$. The anomaly score for the entire spectrogram $X$ can be computed by averaging (mean-pooling) the predictions from the smaller patches:\\
\begin{equation}
    \label{eq:F}
    \mathcal{F}(X) = \frac{1}{n}\sum_{i=1}^{n}\mathcal{F} \circ f(x_i)
\end{equation}
Since we observed that acoustic anomalies of factory machinery can often be spotted visually (see Figure \ref{fig:spectrograms}), we claim that a NN pretrained on the task of image recognition can extract meaningful features that help to distinguish between normal and anomalous operation.
The filters of these networks were shown~\cite{Olah2017,Olah2020} to having learned to recognize colors, contrast, shapes (e.g. lines, edges), objects and textures. Leveraging pretrained NNs is commonly referred to as transfer learning.\\
Note that the simple summation in Equation \ref{eq:F} neglects the temporal dependency between the patches. In our case the signals we study are considerably less complex than e.g. speech or music and, to some extend, exhibit stationary patterns. Thus, we argue that introducing recurrence has only minor benefits at the cost of increased complexity.

%% file: sections/dataset.tex
\section{Dataset}
\label{sec:dataset}
\begin{figure*}
\centering
  \includegraphics[width=1\linewidth]{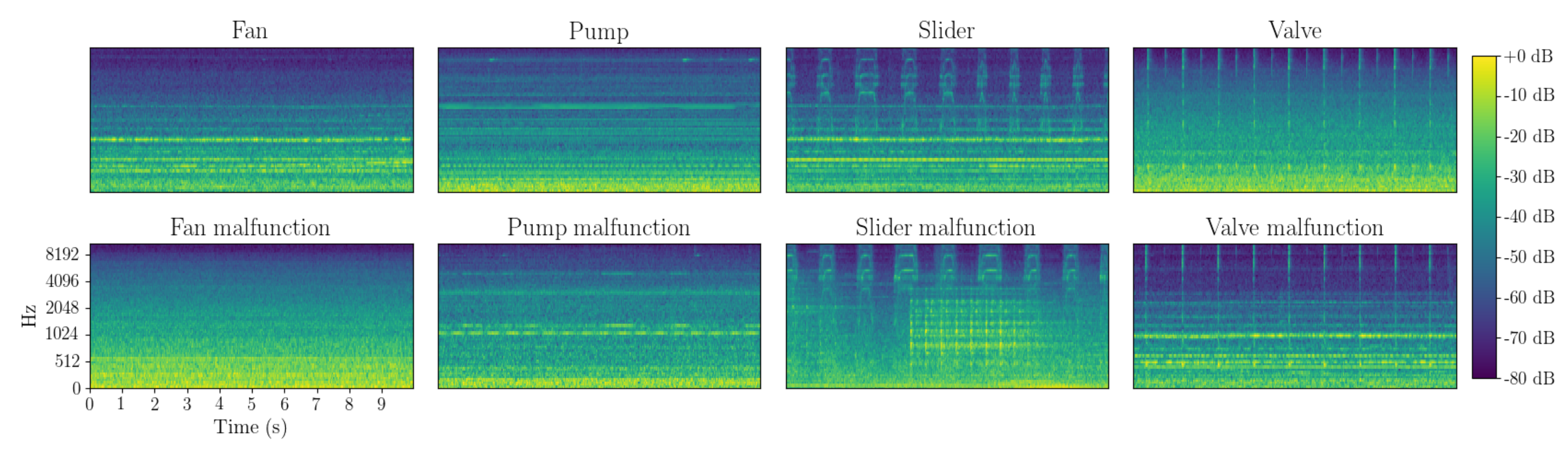}
  \caption{Mel-spectrograms of recordings from normal (top row) and anomalous operation (bottom row) across all machine types in the \textit{MIMII} dataset. Since anomalies can often be spotted visually in this representation, using image classification models is reasonable.}
  \label{fig:spectrograms}
\end{figure*}

In our experiments, we use the recently introduced \textit{MIMII} dataset~\cite{Purohit2019}. It consists of recordings from four industrial machine types (fans, pumps, slide rails and valves) under normal and anomalous operation. For each machine type, four datasets exist, each representing a different product model. Note that anomalous recordings exhibit various scenarios such as leakage, clogging, voltage change, a loose belt or no grease. In addition, background noise recorded in real-world factories was added to each recording according to a certain Signal-to-Noise-Ratio (SNR). In our analysis, we use sounds with a SNR of $-6$dB. We argue that this is very close to the practical use as it is unpreventable that microphones monitoring machines will also capture background noises in a factory environment.
Each single-channel recording is $10$ seconds long and has a sampling rate of $16$kHz. Figure~\ref{fig:spectrograms} depicts Mel-spectrograms of normal and anomalous sounds for all machine types.

%% file: sections/experiments.tex
\section{Experiments}
\label{sec:experiments}
To study the efficacy of image transfer learning for acoustic anomaly detection, we first compute the Mel-Spectrograms for all recordings in the dataset using $64$ Mel-bands, a hanning window of $1024$ and a hop length of $256$. Afterwards, we extract $64 \times 64$ Mel-spectrogram patches ($\approx$ 2s) in a sliding window fashion with an offset of $32$ ($\approx$ 1s) across the time axis and convert them to RGB-images utilizing the \textit{viridis} color-map\footnote{We have found the choice of colormap to be neglectable in terms of performance.}. Subsequently, images are up-scaled ($224\times224$) and standardized using the values obtained from ImageNet to match the domain of the feature extractor $f$. Note that due to our choice of the size of Mel-spectrogram patches, the original aspect-ratio remains unaltered, countering potential information loss. Then, we extract a feature vector for each patch by using various NNs that were pretrained on ImageNet and apply standardization. Finally, we train multiple anomaly detection models on these features.
During training, we randomly exclude $150$ samples, each with a length of $10s$, from the normal data for testing. The same amount of anomalous operation data is randomly added to the test set. A decision for each sample is made using mean pooling, as discussed in Section \ref{sec:proposed_approach}. The whole process is repeated $5$ times with $5$ different seeds and the average \textit{Area Under the Receiver Operating Characteristic Curve} (AUC) is used to report performance.

\subsection{Pretrained Feature Extractors}
Convolutional Neural Networks (CNNs) are known to perform well on two dimensional data input with spatial relations.
Hence, we repurpose the following classifiers, pretrained on ImageNet \cite{imagenet2009} for feature extraction:\\
\textit{Alexnetv3}~\cite{Krizhevsky2012} is a two stream network architecture involving convolutions (kernels: $11 \times 11$, $5 \times 5$ and $3 \times 3$) and max pooling followed by two fully connected layers. We use the activations from the penultimate layer, resulting in a $4096$ dimensional vector.\\
\textit{ResNet18}~\cite{He2016} was designed to counter the problem of diminishing returns when network depth increases. The architecture consists of multiple residual blocks.
16 + 2 layers (initial convolution and max-pooling, followed by 8 convolutional residual blocks) with increasing convolutional filter sizes lead to a single average pooling operation.
We use the $512$ activations thereafter for training.\\
\textit{ResNet34}~\cite{He2016} adheres to the same principles as \textit{ResNet18} at an increased depth of 32 + 2 layers.\\
\textit{SqueezeNet}~\cite{Iandola2016} was designed to use as few parameters as possible ($50$ times fewer than AlexNet) while still providing comparable classification accuracy. This is achieved with the help of \textit{Fire} layers equipped with \textit{squeeze} ($1\times1$) and \textit{expand} ($3\times3$) modules. We apply $2 \times 2$ average pooling to the final feature-map before the classifier to extract a $2048$-dimensional feature vector.

\subsection{Anomaly Detection Models}
We compare six well established anomaly detection algorithms:\\
The \textit{Isolation Forest (IF)}~\cite{Liu2008} is based upon the assumption that anomalies lie in sparse region in feature space and are therefore easier to isolate. Features are randomly partitioned and the average path length across multiple trees is used as the normality score. The number of trees in the forest is set to $128$.\\
A \textit{Gaussian Mixture Model (GMM)} fits a mixture of Gaussians on to the observed features. The log-probability of a feature vector under the trained GMM is used as the normality score. Parameters are estimated via expectation-maximization. We use $80$ mixture components with diagonal covariance matrix initialized using k-means. The iteration limit is set to $150$.\\
The \textit{Bayesian Gaussian Mixture Model (B-GMM)} is trained via variational inference and places prior distributions over the parameters. In many cases, it is less dependent on the specified number of mixtures. In our setting, this might be advantageous as this quantity is hard to determine due to the lack of anomalous data for validation. We use the same parameters as for the GMM.\\
A \textit{One-Class Support Vector Machine (OC-SVM)}~\cite{Scholkopf2000} aims to find the maximum margin hyperplane that best separates the data from the center. 
As $\nu$ (approximate ratio of outliers) must be $>0$, we set $\nu=10^{-4}$ since the training data consists of normal data only.\\
\textit{Kernel Density Estimation (KDE)} is a non-parametric density estimation algorithm that centers a predefined kernel with some  bandwidth over each datapoint and increases the density around this point. Areas with many datapoints will therefore have a higher density than those with only a few. We use a gaussian kernel with a bandwidth of $0.1$. The density at a datapoint is used as normality score.\\
A \textit{Deep Convolutional Autoencoder (DCAE)} reconstructs its own input, in this case the Mel-spectrogram images. We use a LeNet style, three layer convolutional encoder architecture with $32, 64$ and $128$ output channels, a kernel size of $5$, Exponential Linear Unit (ELU)~\cite{Clevert2015} activation functions, batch normalization~\cite{ioffe2015} and a $128$-dimensional bottleneck (LeNet-AE). Moreover, we also consider a simpler encoder architecture with $12, 24$ and $48$ output channels, ReLU~\cite{Nair2010} activation functions and a kernel size of $4$ (Small-DCAE). 
The decoders mirror the encoders using de-convolutional layers. For optimization, we use Adam with learning rate $=10^{-4}$, batch size $=128$ and train for $80$ epochs. The mean squared error between the original image and the reconstruction is used as the loss function and anomaly score. 
\input{sections/results_table}

%% file: sections/results_table.tex
\begin{table*}[htb]

\caption{Anomaly detection results for all machine types and machine IDs (M0, M2, M4 and M6). The best performing model (read vertically) is written in bold and colored in green, the second best is underlined and colored in yellow. Each entry represents the average AUC across five seeds.}
\label{tab:results}
\centering
\scriptsize
\begingroup
\setlength{\tabcolsep}{5.0pt} 
\renewcommand{\arraystretch}{1.15} 
\begin{tabular}{cl|llll|llll|llll|llll}
\multicolumn{1}{l}{}         &                        & \multicolumn{4}{c|}{Fan}                                                                                                                                      & \multicolumn{4}{c|}{Pump}                                                                                                                                                            & \multicolumn{4}{c|}{Slider}                                                                                                                                   & \multicolumn{4}{c}{Valve}                                                                                                                                     \\
\multicolumn{1}{l}{}         &                        & M0                                    & M2                                    & M4                                    & M6                                    & M0                                    & M2                                    & M4                                    & M6                                                           & M0                                    & M2                                    & M4                                    & M6                                    & M0                                    & M2                                    & M4                                    & M6                                    \\ \hline
                             & GMM                    & 57.7                                  & 61.7                                  & 53.9                                  & \cellcolor[HTML]{5AC865}\textbf{94.5} & \cellcolor[HTML]{F9E721}{\ul 84.1}    & \cellcolor[HTML]{F9E721}{\ul 70.8}    & 81.6                                  & 66.0                                                         & 98.3                                  & 80.9                                  & 61.4                                  & 57.5                                  & 60.2                                  & 69.2                                  & 59.9                                  & 53.5                                  \\
                             & B-GMM                  & 50.9                                  & 61.4                                  & 47.7                                  & 82.2                                  & 71.8                                  & 60.2                                  & 73.4                                  & 53.3                                                         & 83.2                                  & 65.0                                  & 50.0                                  & 57.0                                  & 55.2                                  & 62.7                                  & 51.4                                  & 48.3                                  \\
                             & IF                     & 53.1                                  & 59.7                                  & 48.9                                  & 84.6                                  & 75.9                                  & 62.4                                  & 75.0                                  & 55.9                                                         & 89.4                                  & 69.0                                  & 51.9                                  & 56.2                                  & 50.1                                  & 63.4                                  & 53.3                                  & 49.8                                  \\
                             & KDE                    & 55.7                                  & 59.1                                  & 50.5                                  & 90.3                                  & 76.4                                  & 65.9                                  & 74.8                                  & 61.0                                                         & 97.8                                  & 79.3                                  & 59.7                                  & 55.0                                  & 54.6                                  & 64.4                                  & 57.1                                  & 51.4                                  \\
\multirow{-5}{*}{AlexNet}    & OC-SVM                 & 51.0                                  & \cellcolor[HTML]{5AC865}\textbf{73.1} & \cellcolor[HTML]{5AC865}\textbf{59.7} & 93.2                                  & 77.5                                  & 56.4                                  & 81.1                                  & 60.1                                                         & 96.2                                  & 81.4                                  & 53.6                                  & 56.5                                  & 61.6                                  & 73.6                                  & 48.3                                  & 48.9                                  \\ \hline
                             & GMM                    & \cellcolor[HTML]{5AC865}\textbf{62.6} & 64.1                                  & \cellcolor[HTML]{F9E721}{\ul 59.3}    & \cellcolor[HTML]{F9E721}{\ul 94.4}    & \cellcolor[HTML]{5AC865}\textbf{84.5} & \cellcolor[HTML]{5AC865}\textbf{71.3} & 84.0                                  & \cellcolor[HTML]{F9E721}{\color[HTML]{000000} {\ul 68.3}}    & 99.1                                  & \cellcolor[HTML]{F9E721}{\ul 85.8}    & 68.8                                  & 65.6                                  & 58.3                                  & 73.3                                  & 60.2                                  & 56.9                                  \\
                             & B-GMM                  & \cellcolor[HTML]{F9E721}{\ul 59.2}    & 60.5                                  & 54.8                                  & 91.0                                  & 79.1                                  & 69.7                                  & 79.4                                  & 59.5                                                         & 98.3                                  & 77.7                                  & 61.4                                  & 61.2                                  & 70.1                                  & 71.7                                  & 56.1                                  & 50.3                                  \\
                             & IF                     & 58.0                                  & 60.5                                  & 55.3                                  & 86.5                                  & 70.8                                  & 59.0                                  & 77.3                                  & 54.6                                                         & 97.7                                  & 72.7                                  & 60.6                                  & 61.2                                  & 56.5                                  & 69.8                                  & 58.2                                  & 47.5                                  \\
                             & KDE                    & 57.9                                  & 59.1                                  & 55.6                                  & 85.9                                  & 76.6                                  & 56.5                                  & 76.7                                  & 62.2                                                         & 98.1                                  & 77.0                                  & 61.2                                  & 60.9                                  & 57.6                                  & 62.9                                  & 56.8                                  & 49.7                                  \\
\multirow{-5}{*}{ResNet18}   & OC-SVM                 & 55.0                                  & \cellcolor[HTML]{F9E721}{\ul 68.8}                            & 57.4                                  & 87.7                                  & 71.6                                  & 55.2                                  & 78.6                                  & 60.6                                                         & 96.7                                  & 79.6                                  & 69.3                                  & 66.2                                  & 61.1                                  & 76.1                                  & 56.8                                  & 43.1                                  \\ \hline
                             & GMM                    & 58.7                                  & 65.6                                  & 57.0                                  & 90.9                                  & 78.4                                  & 66.8                                  & \cellcolor[HTML]{F9E721}{\ul 87.9}    & 63.2                                                         & \cellcolor[HTML]{5AC865}\textbf{99.6} & \cellcolor[HTML]{5AC865}\textbf{90.4} & \cellcolor[HTML]{5AC865}\textbf{82.5} & 69.1                                  & \cellcolor[HTML]{F9E721}{\ul 73.0}    & \cellcolor[HTML]{5AC865}\textbf{79.1} & 60.1                                  & \cellcolor[HTML]{5AC865}\textbf{61.9} \\
                             & B-GMM                  & 55.7                                  & 61.8                                  & 52.3                                  & 85.8                                  & 71.5                                  & 61.1                                  & 84.5                                  & 55.2                                                         & \cellcolor[HTML]{F9E721}{\ul 99.2}    & 85.4                                  & \cellcolor[HTML]{F9E721}{\ul 72.3}    & 63.6                                  & 70.8                                  & 76.2                                  & 59.3                                  & 57.9                                  \\
                             & IF                     & 53.9                                  & 62.0                                  & 49.9                                  & 82.2                                  & 52.3                                  & 48.3                                  & 79.3                                  & 49.4                                                         & 98.6                                  & 83.1                                  & 69.5                                  & 60.2                                  & 65.9                                  & 71.2                                  & \cellcolor[HTML]{F9E721}{\ul 60.3}    & 54.0                                  \\
                             & KDE                    & 55.0                                  & 62.6                                  & 52.3                                  & 83.1                                  & 62.0                                  & 51.8                                  & 82.8                                  & 58.3                                                         & 99.0                                  & 84.0                                  & 68.2                                  & 62.2                                  & 67.5                                  & 71.9                                  & 53.9                                  & 58.2                                  \\
\multirow{-5}{*}{ResNet34}   & OC-SVM                 & 50.1                                  & 67.4                                  & 57.5                                  & 83.0                                  & 64.9                                  & 51.5                                  & 81.2                                  & 60.2                                                         & 96.8                                  & 85.0                                  & 71.4                                  & 64.3                                  & \cellcolor[HTML]{5AC865}\textbf{75.6} & \cellcolor[HTML]{F9E721}{\ul 77.8}    & \cellcolor[HTML]{5AC865}\textbf{64.3} & 53.1                                  \\ \hline
                             & GMM                    & 56.1                                  & 60.4                                  & 49.4                                  & 83.4                                  & 72.1                                  & 46.4                                  & 87.6                                  & 60.8                                                         & 96.7                                  & 76.8                                  & 52.1                                  & 62.9                                  & 62.8                                  & 75.3                                  & 53.3                                  & 57.3                                  \\
                             & B-GMM                  & 54.4                                  & 59.8                                  & 47.0                                  & 84.5                                  & 72.3                                  & 48.2                                  & 86.2                                  & 69.0                                                         & 95.0                                  & 78.8                                  & 55.8                                  & 65.0                                  & 63.8                                  & 74.0                                  & 52.4                                  & 56.8                                  \\
                             & IF                     & 53.2                                  & 64.0                                  & 44.8                                  & 84.6                                  & 76.1                                  & 45.5                                  & 85.3                                  & 60.2                                                         & 98.9                                  & 78.2                                  & 53.1                                  & \cellcolor[HTML]{F9E721}{\ul 70.6}    & 56.6                                  & 68.7                                  & 51.5                                  & 56.6                                  \\
                             & KDE                    & 54.4                                  & 60.5                                  & 47.0                                  & 84.3                                  & 74.5                                  & 45.2                                  & 86.5                                  & 61.4                                                         & 98.7                                  & 80.8                                  & 56.4                                  & 69.2                                  & 65.0                                  & 74.5                                  & 52.8                                  & 57.7                                  \\
\multirow{-5}{*}{SqueezeNet} & OC-SVM                 & 55.6                                  & 64.8                                  & 46.2                                  & 86.7                                  & 78.8                                  & 49.4                                  & \cellcolor[HTML]{5AC865}\textbf{88.4} & 62.3                                                         & 99.2                                  & 81.5                                  & 59.4                                  & \cellcolor[HTML]{5AC865}\textbf{71.6} & 69.0                                  & 71.3                                  & 53.1                                  & \cellcolor[HTML]{F9E721}{\ul 58.2}    \\ \hline
LeNet-DCAE                   & \multicolumn{1}{c|}{-} & 49.1                                  & 57.0                                  & 53.2                                  & 66.9                                  & 65.3                                  & 54.4                                  & 76.0                                  & \cellcolor[HTML]{FFFFFF}66.6                                 & 95.9                                  & 70.4                                  & 56.2                                  & 50.6                                  & 42.3                                  & 55.6                                  & 51.2                                  & 45.5                                  \\
Small-DCAE                   & \multicolumn{1}{c|}{-} & 48.3                                  & 54.1                                  & 49.3                                  & 63.7                                  & 69.9                                  & 52.9                                  & 73.1                                  & \cellcolor[HTML]{5AC865}{\color[HTML]{000000} \textbf{69.2}} & 95.3                                  & 68.4                                  & 55.7                                  & 53.3                                  & 36.6                                  & 57.2                                  & 51.2                                  & 45.4                                 
\end{tabular}
\endgroup
\end{table*}

%% file: sections/results.tex
\section{Results}
\label{sec:results}
In this section, we discuss the key findings of the results depicted in Table \ref{tab:results}. Note that these findings only refer to the setting introduced in the prior chapters.\\
\textbf{1) Image Transfer Learning is more effective for detecting anomalous machine sounds than autoencoders trained from scratch.}\\
Autoencoders outperform the models based on image transfer learning only in a single setting (Small-DCAE on Pump-M6). In the majority of the cases, LeNet-DCAE yields better results than Small-DCAE. Mostly, the DCAEs do not even come close to their competitors, which supports our hypothesis that the features extracted by learned filters from pretrained image classification models are better suited for detecting subtle anomalies. Further, reconstruction based anomaly detection is based upon a proxy task rather than modeling the task explicitly.\\
\textbf{2) ResNet architectures are superior feature extractors.}\\
To compare the feature extractors, we count the scenarios in that a specific feature extractor combined with different anomaly detection models yields the highest or the second highest score and create tuples of the form $(1^{\text{st}}, 2^{\text{nd}})$. As depicted in Table \ref{tab:results}, there are $16$ distinct evaluation settings in which either the highest or the second highest score can be achieved. Ranked from best to worst, we get the following results: ResNet34 $(7, 6)$, ResNet18 $(3, 5)$, AlexNet $(3, 2)$, SqueezeNet $(2, 2)$ and Autoencoders $(1, 0)$. A clear superiority of ResNet based feature feature extractors can be observed. Interestingly, these are also the models with a lower classification error on ImageNet compared to SqueezeNet and AlexNet. These results are consistent with a recent finding that there is strong correlation between ImageNet top-1 accuracy and transfer learning accuracy~\cite{Kornblith2019}.
Another important observation is that ResNet34's good performance almost exclusively stems from top performance on sliders and valves. The Mel-spectrogram images from these machines have more fine granular variations than those from fans and pumps which show a more stationary allocation of frequency bands. We assume that ResNet34 extracts features on a more detailed level which can explain inferior performance on fan and pump data. Generally, we have found SqueezeNet to be the least reliable feature extractor. Note that these findings also hold when all feature vectors are reduced to the same dimensionality using Principle Components Analysis (PCA).\\
\textbf{3) GMM and OC-SVM yield the best performance.}\\
To compare the anomaly detection models, we count the scenarios in that a specific anomaly detection model combined with different feature extractors achieves the best or second best result. Employing the same ranking strategy as above, the results are as follows: GMM $(9, 8)$, OC-SVM $(6, 2)$, Autoencoders $(1, 0)$, B-GMM $(0, 3)$, IF $(0, 2)$, KDE $(0, 0)$. Clearly, GMM and OC-SVM outperform all other models by a large margin. Together, they account for $15/16$ of the best performing models and $10/16$ of second best performing models. Although GMM and B-GMM are both based on the same theoretical assumptions, B-GMM produces inferior results. We suspect the weight priors to potentially be too restrictive.\\
\textbf{4) Results are highly dependent on the machine type and the machine model.}\\
The model performing best on valves has an average AUC of $79.1$. This is low compared to the other machine types as these always have at least one scenario with an average AUC $> 80$. Moreover, the highest achieved score varies considerably across all machine types. This indicates that some machine types are more suited for our approach (pumps, sliders) than others (fans, valves). More importantly, a significant variance between different machine IDs (M0 - M6) can be observed. Results on fans make this problem most evident. While M0, M2 and M4 have average scores of $62.6, 73.1$ and $59.7$, M6 achieves an average of $94.5$. M6 improves upon M4 at $\approx 30\%$. This suggests that anomalous sound patterns are vastly different (more or less subtle) even for different models of the same machine type. Future approaches should take this into account.\\

%% file: sections/conclusion.tex
\section{Conclusion}
\label{sec:conclusion}
In this work, we thoroughly studied acoustic anomaly detection for machine sounds.
For feature extraction, we used readily available neural networks that were pretrained to classify ImageNet images.\\
We then used these features to train five different anomaly detection models. Results indicate that features extracted with ResNet based architectures in combination with a GMM or an OC-SVM yield the best average AUC.\\
Moreover, we confirmed our hypothesis that the image based features are general purpose and consequently also yield Competitive acoustic anomaly detection results.\\
Future work could investigate upon further ensemble approaches and other feature extraction architectures~\cite{Kawaguchi2019,Pons2019,Howard2017,Huang2016}. In addition, our approach might benefit from techniques that reduce background noise~\cite{Zhang2018} or enable decisions over a longer time-horizon~\cite{Xie2019}.
One might also try to use pretrained feature extractors from other, more related domains such as music or environmental sounds.